\documentclass{article}
\usepackage{amsbsy,amssymb}
\newcommand{\be}{\begin{equation}}
\newcommand{\ee}{\end{equation}}
\newcommand{\Tr}{{\rm Tr}}
\def\bea{\begin{eqnarray}}
\def\eea{\end{eqnarray}}
\def\bean{\begin{eqnarray*}}
\def\eean{\end{eqnarray*}}

\newcommand{\barr}{\begin{array}}
\newcommand{\earr}{\end{array}}

\newcommand{\bed}{\begin{displaymath}}
\newcommand{\eed}{\end{displaymath}}
\newcommand{\bal}{\begin{array}{ll}}
\newcommand{\eal}{\end{array}}

\def\bvec#1{\raise1.5ex\hbox{$\rightarrow$}\mkern-16.5mu #1}

\def\m#1{\mathcal#1}

\begin{document}
\title{\hfill ~\\[-30mm]
       \hfill\mbox{\small UFIFT-HEP-08-4}\\[30mm]
       \textbf{Quintics with  Finite Simple Symmetries }} 
\date{}
\author{\\Christoph Luhn,\footnote{E-mail: {\tt luhn@phys.ufl.edu}}~~
        Pierre Ramond\footnote{E-mail: {\tt ramond@phys.ufl.edu}}\\ \\
  \emph{\small{}Institute for Fundamental Theory, Department of Physics,}\\
  \emph{\small University of Florida, Gainesville, FL 32611, USA}}

\maketitle

\begin{abstract}
\noindent We construct all quintic invariants in five variables with {\em
  simple} Non-Abelian finite symmetry groups. These define Calabi-Yau
three-folds which are left invariant by the action of  $\m A_5$, $\m
A_6$ or $\mathcal{PSL}_2(11)$.  
\end{abstract}
\thispagestyle{empty}
\vfill
\newpage
\setcounter{page}{1}

\section{Introduction}
There has long been interest in the study of hypersurfaces defined by a
quintic polynomial in five variables. Quintic polynomials in five complex
variables characterize a class of Calabi-Yau three-folds over $CP^4$, see
Ref.~\cite{Doran:2007jw}. A special subclass of such manifolds is that where
the polynomial is invariant under discrete groups.   

The identification of those hypersurfaces which are left invariant under the
action of Non-Abelian finite groups proceeds in two steps.  First,  identify Non-Abelian finite groups with five-dimensional irreducible representations. Second,  single out those which can sustain a quintic invariant. 

Unlike the case of two, three and four
variables~\cite{Miller,Blichfeldt,Fairbairn,Hanany:1999sp}, there is no
catalog of all finite groups with five-dimensional representations. However it
is an easy matter to find the {\em simple} groups which have such
representations. The ATLAS~\cite{atlas} lists just a handful of {\em simple}
groups: 

\begin{itemize}

\item The $60$ element symmetry group of the isocahedron and dodecahedron, aka the Icosahedral group $\mathcal I=\m A_5= \mathcal{PSL}_2(5)$ with one {\em real} five-dimensional representation.

\item The Alternating group of even permutations on six objects, $\m A_6$ with two inequivalent {\em  real }five-dimensional irreps. 

\item The group of projective linear transformations of unit determinant over the Galois field of order $11$, $\mathcal{PSL}_2(11)$ with $660$ elements has one {\em complex} five-dimensional irrep ${\bf 5}$ and its conjugate ${\bf\bar 5}$.

\item The group of $(4\times 4)$ unitary matrices over the Galois field of
  order two, $U_4(2)=Sp_4(3)$ with $25920$ elements also has one {\it complex}
  irrep ${\bf 5}$ and the~${\bf\bar 5}$. 

\end{itemize}
It is known that the real $\m A_5$ quintet has {\em two} quintic
invariants, and {\it two} cubic invariants. Their explicit construction can be found in the beautiful paper of Cummins and Patera~\cite{Cummins:1988dr}, which serves as a starting point for our investigations.  Our new results are simply stated:

\begin{itemize}

\item We identify and construct one quintic and one cubic $\m A_6$ invariant from its quintet representation. 

\item We find only one $\mathcal{PSL}_2(11)$  quintic invariant on five complex variables, and give its explicit form. 

\item We show that the   $U_4(2)=Sp_4(3)$ complex quintet  does {\em not} have a quintic invariant, by constructing its Molien polynomial (see below).

\end{itemize}
We present these results in the hope that they will motivate the study of these special hypersurfaces, in particular the unique $\mathcal{PSL}_2(11)$ Calabi-Yau manifold.
\subsection{\label{molien}Finite Group Invariants}
The construction of finite group invariants \cite{Sloane,PateraSharpWinter} is a complicated but tractable problem. Finite groups have, up to conjugation,
one irreducible representation whose Kronecker products generate all other irreducible representations. One can in this way find all invariants made out
of {\it one} such irreducible representation. The procedure can be duplicated for any other irreducible representation. The same technique can be continued in principle to the building of invariants constructed out of several irreducible representations, although it becomes quite unwieldy.  

Consider a discrete group $\m G$ of order $N$, with $l$ irreducible representations $\bf{r_a}$, with $\bf{r_1}={\bf 1}$ the singlet. The $k-$fold  symmetric Kronecker product

\be
\underbrace{({\bf{r_a}\otimes \bf{r_a}\otimes\cdots \otimes
    \bf{r_a}})_s}^{}_k~=~\sum_{b} D^{[k]}({\bf r_b}; \bf{r_a})\,\bf{r_b^{}}\ ,
\ee
is a reducible representation, the sum of the group's irreps. The integer
coefficient $D^{[k]}({\bf r_b}; \bf{r_a})$ is the multiplicity of the
$\bf{r_b}$ representation in the product. These coefficients are determined
from Molien's remarkable generating function~\cite{Molien}

\be
F({\bf r_b};{\bf r_a};\,\lambda)~=~\frac{1}{N}\sum _{i=1}^l\,n_i^{}\frac{{\overline\chi}_{ i}^{[{\bf r_b}]}}{\det(1-\lambda A^{[{\bf r_a}]}_i)}\ ,
\ee
where $i$ labels the class $C_i$ of $\m G$, with $n_i$  elements. In the $i$th  class, $A^{[{\bf r_a}]}_i$ is any group element in the $\bf r_a$ representation, and $\chi_i^{[{\bf r_b}]}$ is the character of the $\bf r_b$ representation. Its power series 

\be
F({\bf r_b};{\bf r_a};\,\lambda)~=~\sum_k\,D^{[k]}({\bf r_b};
\bf{r_a})\,\lambda^k_{}\ ,
\ee
yields the desired coefficients.  The Molien function is an important tool in determining all invariants constructed out of one irreducible representation 

\be
F({\bf1};{\bf r};\,\lambda)~=~\frac{1}{N}\sum _{i=1}^l\,\frac{n_i^{}}{\det(1-\lambda A^{[{\bf r}]}_i)}~=~\sum_{k=0}^\infty c^{}_k\,\lambda^k_{}\ ,
\ee
where $c_k$ denotes the number of invariants of order $k$.  The evaluation of the finite sum over the classes yields an
expression which can always be written in the form

\be
F({\bf1};{\bf r};\,\lambda)~=~\frac{(1+\sum d_k^{}\,\lambda^k_{})}{(1-\lambda^{a_1}_{})^{n_1}_{}(1-\lambda^{a_2}_{})^{n_2}_{}\cdots}\ ,
\ee
where the numerator is a {\em finite } polynomial in $\lambda$, and the $d_k$, $a_k$ and $n_k$ are positive integers. Expanding each factor in the denominator yields $d_k$ invariants of order $k$, as well as $n_k$ invariants of order $a_k$, etc... . This infinite number of  invariants can be expressed as products of  a finite set of {\em basic} invariants, some of which satisfy non-linear relations called {\em syzygies} among themselves.

In some simple cases, the invariants of order $a_1$, $a_2$, etc..., as indicated by the denominator in the Molien series do not satisfy syzygies among themselves. They are called {\em free invariants} by the authors of reference [7]. The  powers in the numerator refer to  {\em constrained} invariants which satisfy syzygies with the free invariants. Unfortunately, this neat distinction among invariants is of limited validity. 

To see how this works, consider the Tetrahedral group $\m A_4$, which has four irreps, ${\bf 1},\,{\bf 1_1},\,{\bf \bar 1_1},$ and ${\bf 3}$. 
Using its character table, it is an easy matter to compute the generating function for each. For the singlet irrep, we have of course

$$F({\bf 1};{\bf 1};\,\lambda)~=~\frac{1}{1-\lambda}\ ,$$
corresponding to the trivial invariant. The situation is a bit less trivial for the other one-dimensional representations,

$$F({\bf 1};{\bf 1_1};\,\lambda)~=~F({\bf 1};{\bf \bar 1_1};\,\lambda)~=~\frac{1}{1-\lambda^3_{}}\ ,$$
which means that there is one invariant of cubic order. If $z$ spans the irrep, then $z^3$ is an invariant. For the triplet, the situation becomes more complicated. From

$$F({\bf 1};{\bf 3};\,\lambda)~=~\frac{1+\lambda^6_{}}{(1-\lambda^2_{})(1-\lambda^3_{})(1-\lambda^4_{})}\ ,$$
we deduce that there are three free invariants, of order $2,3$ and $4$, and one sixth order constrained invariant. Assume that the triplet irrep is spanned by the three real coordinates $x_i^{}$, $i=1,2,3$. The quadratic invariant is the same as for $SO(3)$

$$I^{[2]}_{{\bf 3}}~=~(x^2_1+x^2_2+x^2_3)\ .$$
The cubic and quartic invariants are given by

$$I^{[3]}_{{\bf 3}}~=~x^{}_1x^{}_2x^{}_3\ ,\qquad I^{[4]}_{{\bf 3}}~=~x^4_1+x^4_2+x^4_3\ .$$
The constrained sixth-order invariant is 

$$
E^{[6]}_{{\bf 3}}~=~(x^2_1-x^2_2)(x^2_2-x^2_3)(x^2_3-x^2_1)\ .$$
The syzygy, too long to write here, expresses the square of the sixth-order invariant in terms of polynomials in the free invariants.

\section{\label{A5sec}Quintics from ${\boldsymbol{\mathcal A_5}}$} 
We begin our study with the Icosahedral group, $\m A_5=\mathcal{PSL}_2(5)$. It is generated by two elements with presentation 

\be
<A,B\,|\,A^2=B^3=(AB)^5=1\,>\ ,
\ee
and character table
\vskip .3cm
{\footnotesize{\begin{center}
\begin{tabular}{c|ccccc}
&   \\
{\Large$ \m A_5$}~~~~& $C_1$&\hfill $12C^{[5]}_2$& \hfill$12C^{[5]}_3$&\hfill $15C^{[2]}_4$&\hfill $20C^{[3]}_5$\\
 &&&&&   \\
\hline    
 &&&&&   \\
$\chi^{[\bf 1]}_{}$~~& $1$&\hfill $1$&\hfill  $1$&\hfill  $1$&\hfill $1$\\
 &&&&&   \\
$\chi^{[\bf 3_1]}_{}$~~& $3$&\hfill  $-b^{}_5$&\hfill  $-\tilde b_5^{}$&\hfill  $-1$&\hfill $0$ \\
 &&&&&   \\
$\chi^{[\bf 3_2]}_{}$~~& $3$&\hfill  $-\tilde b_5^{}$&\hfill  $-b_5^{}$&\hfill  $-1$&\hfill $0$\\
 &&&&&   \\
$\chi^{[\bf 4]}_{}$~~& $4$&\hfill  $-1$&\hfill  $-1$&\hfill  $0$&\hfill $1$\\
 &&&&&   \\
$\chi^{[\bf 5]}_{}$~~& $5$& \hfill $0$&\hfill  $0$&\hfill  $1$&\hfill $-1$\\
\end{tabular}\end{center}
}}
\vskip .5cm
\noindent where

\be 
b^{}_5~=~\frac{1}{2}(-1+\sqrt{5})\ ,\qquad \tilde b^{}_5~=~\frac{1}{2}(-1-\sqrt{5})\ .
\ee
In the Cummins-Patera basis~\cite{Cummins:1988dr}, the quintic representation generators are given by   

\be
A\;= \pmatrix{
\hfill 1&\hfill 0&\hfill 0&\hfill 0&\hfill 0\cr 
\hfill 0&\hfill -1&\hfill 0&\hfill 0&\hfill 0\cr
\hfill 0&\hfill 0&\hfill -1&\hfill 0&\hfill 0\cr
\hfill 0&\hfill 0&\hfill 0&\hfill 1&\hfill 0\cr
\hfill 0&\hfill 0&\hfill 0&\hfill 0&\hfill 1} ;
\quad
B\;=\;-\frac{1}{2}\pmatrix{1&0&1&-\omega^2&-\omega\cr
0&1&1&-1&-1\cr
-1&-1&0&\omega&\omega^2\cr
\omega&1&\omega^2&0&-\omega\cr
\omega^2&1&\omega&-\omega^2&0} , \label{AandB}
\ee
where $\omega=\exp(2i\pi/3)$ is the cubic root of unity, 

\be
\omega^3=1\ ,\quad 1+\omega+\omega^2~=~0 \ ,\quad \omega=-\frac{1}{2}+\frac{i}{2}\sqrt{3}\ .
\ee
These matrices act on the column vector

\be
\pmatrix{x^{}_1\cr x^{}_2\cr x^{}_3\cr z\cr \overline z} , \qquad
z=\frac{1}{\sqrt{2}}(x^{}_4+ix^{}_5)\ ,
\ee
where the $x_i$ are real.  The Molien generating function for the quintet is

$$
\frac{1+\lambda^5_{}+2\lambda^6_{}+\lambda^7_{}+\lambda^{12}_{}}{(1-\lambda_{}^2)(1-\lambda_{}^3)^2(1-\lambda_{}^4)(1-\lambda_{}^5)}\
.$$ From the powers of $\lambda$ in the denominator, we see that from the quintet we can form five free invariants: one quadratic, two cubic, one quartic, and one quintic.  
The powers in the numerator indicate several constrained invariants, another quintic invariant, as well as two of  sixth-order, one of seventh-order, and one of twelfth-order. 
  
The quadratic invariant is the obvious one\\[0mm]
\be
I^{[2]}_{\,\bf
  5}~=~x_1^2+x_2^2+x_3^2+x_4^2+x_5^2~=~x_1^2+x_2^2+x_3^2+2z\overline z\ .
\ee
One  cubic invariant\\[-3mm]
\bea
I^{[3]}_{a\,\bf 5}&=&z^3-\overline z^3+3(x^2_1+\omega x^2_2+\omega^2
x^2_3)\overline z-3(x^2_1+\omega^2 x^2_2+\omega x^2_3) z \nonumber \\
&=&-\frac{i}{\sqrt{2}}\left\{x_5^3+3\sqrt{3}\,x^{}_4(x_3^2-x_2^2)+3x^{}_5(2x_1^2-x_2^2-x_3^2-x_4^2)\right\} ,\eea
is pure imaginary, that is odd under complex conjugation\\[-1mm]
$$z~\rightarrow ~\overline z\ ,\quad \overline z~\rightarrow ~ z \ ; \quad\omega~\rightarrow ~\omega^2\ ,\quad \omega^2~\rightarrow ~\omega\ ,$$
or in terms of the vector components\\[-1mm]
$$z~\rightarrow ~\overline z\ ,\quad \overline z~\rightarrow ~ z\ ; \quad x_2~\rightarrow ~x_3\ ,\quad x_3~\rightarrow ~x_2\ .$$
The second cubic \\[-3mm]
\bea
I^{[3]}_{s\,\bf 5}&=&z^3+\overline z^3+(x^2_1+\omega x^2_2+\omega^2 x^2_3)\overline z+(x^2_1+\omega^2 x^2_2+\omega x^2_3) z-4x^{}_1x^{}_2x^{}_3 \nonumber\\
&=&\frac{1}{\sqrt{2}}\left\{\sqrt{3}\,x^{}_5(x_2^2-x_3^2)+x^{}_4(2x_1^2-x_2^2-x_3^2+x_4^2-3x_5^2)-4\sqrt{2}\,x^{}_1x^{}_2x^{}_3\right\} , \nonumber\\[-1mm]\eea
is real. The one  quartic invariant is\\[-3mm]
\bea
I^{[4]}_{\,\bf 5}&=& x^4_1+x^4_2+x^4_3-2\left[(x^2_1+\omega x^2_2+\omega^2 x^2_3)z^2+(x^2_1+\omega^2 x^2_2+\omega x^2_3)\overline z^2\right]+\nonumber\\
&&+4z\overline z(x_1^2+x_2^2+x_3^2+2z\overline z)-6(z\overline z)^2
\nonumber\\&=&\frac{1}{2}\left\{2x_1^4+2x_2^4+2x_3^4+(x_4^2+x_5^2)^2+4\sqrt{3}\,x^{}_4x^{}_5(x^2_2-x^2_3)\right.\nonumber\\
~~~&&+2x_5^2(4 x_1^2+x_2^2+x_3^2)+6x_4^2(x_2^2+x_3^2)\Big\}\ .\eea
There are two quintic invariants. One is odd under conjugation \\[-3mm]
\bean
I^{[5]}_{a\,\bf 5}&=& (x^4_1+\omega x^4_2+\omega^2 x^4_3)\overline z-(x^4_1+\omega^2 x^4_2+\omega x^4_3) z\\&& -(2x^2_1+2x^2_2+2x^2_3+2z\overline z)\left[(x^2_1+\omega x^2_2+\omega^2 x^2_3)\overline z-(x^2_1+\omega^2 x^2_2+\omega x^2_3) z\right]\\
&&-z\overline z(z^3-\overline z^3)
\ ,\eean
or explicitly\\[-3mm]
\begin{eqnarray}
I^{[5]}_{a\,{\bf 5}}&=&\frac{i}{\sqrt {2}} \bigg\{\sqrt {3}x_3^{4}x_4- \sqrt {3}x_4^{3}
 x_2^{2}-\sqrt {3} x_2^{4} x_4+ \sqrt {3}x_4^{3} x_3^{2}+
\sqrt {3} \left( {2} x_3^{2} x_4- {2} x_2^{2} x_4 \right)  x_1^{2} \nonumber\\ &&
+x^{}_5\left[\frac{-3}{2} x_4^{4}- x_4^{2} x_2^{2}- x_3^{4}-x_2^{4}-4 x_2^{2} x_3^{2}- x_4^{2}
 x_3^{2}+ \left(  {2}x_4^{2}+ {2} x_3^{2}
+ {2} x_2^{2} \right)  x_1^{2}+ {2} x_1^
{4}\right]
\nonumber\\ &&\left.
+\sqrt {3}x^2_5\left[ x_3^{2} x_4- x_2
^{2} x_4\right]- x^3_5\left[x_2^{2}- {2} x_1^{2}+x_4^{2}+ x_3^{2}\right]
+\frac{1}{2}x^5_5\right\}\ . 
\end{eqnarray}
The ``real'' quintic invariant is given by\\[-3mm]
\bean
I^{[5]}_{s\,\bf 5}&=&(x^4_1+\omega x^4_2+\omega^2 x^4_3)\overline z+(x^4_1+\omega^2 x^4_2+\omega x^4_3) z\\ &&
-(2x_1^2+2x_2^2+2x_3^2-2z\overline z)\left[(x^2_1+\omega x^2_2+\omega^2
  x^2_3)\overline z+(x^2_1+\omega^2 x^2_2+\omega x^2_3) z\right]  \\
&&+16x^{}_1x^{}_2x^{}_3z\overline z-z\overline z(z^3+\overline z^3)\ ,
\eean
or in terms of the five real variables\\[-3mm]
\begin{eqnarray}
I^{[5]}_{s\,{\bf 5}}&=&\frac{1}{\sqrt{2}}\left\{\frac{3}{2} x_4\,x_5^4+\sqrt {3}x_5^3\left[
 x_2^{2}- x_3^{2}\right]
-x_5^2\left[ x_3^{2} x_4-{2}x_1^{2} x_4-
 x_4^{3}-8\,\sqrt{2} x_1\, x_2\, x_3+ x_2^{2} x_4\right]\right.\nonumber\\&&
+\sqrt{3}x^{}_5\left[ x_3^{2}( 2x_1^{2}-x_4^{2}+x_3^2)+x_2^2(x_4^2-x_2^2-2x_1^2)\right]-\frac{1}{2}x_4^{5}+
8\, \sqrt{2} x_1\, x_2\,x_3\, x_4^{2}\nonumber\\ &&
+x^{}_4x^2_2(4x_3^2+x_2^2-2x_1^2-x_4^2)
+x^{}_4(x_3^4-2x_1^4+2x_1^2x_4^2-x_3^2x_4^2-2x_1^2x_3^2)\bigg\}\ .
\end{eqnarray}
The most general quintic invariant can therefore be written as

\be
I^{[5]}_{\m A_5}(\alpha,\beta,\gamma,\delta; \vec x)~=~\alpha\,I^{[5]}_{s\,{\bf
    5}}+\beta\, I^{[5]}_{a\,{\bf 5}}+\left(\gamma\, I^{[3]}_{s\,{\bf 5}}+\delta\, I^{[3]}_{a\,{\bf 5}}\right)\,I^{[2]}_{{\bf 5}}\ ,
\ee
with arbitrary coefficients $\alpha$, $\beta$, $\gamma$, $\delta$, and variables

\[
\vec x~=~(x^{}_1,x^{}_2,x^{}_3,x^{}_4,x^{}_5)\ .
\]
The higher order invariants and their syzygies can be found in compact notation in Ref.~\cite{Cummins:1988dr}.

\section{ ${\boldsymbol{\mathcal A_5}}$ Quartic}
Manifolds defined by quartic polynomials in four  variables generate K3
surfaces. It might be instructive  to use our group methods to catalog all such
cases, and compare it with the classification~\cite{Xiao} of K3 manifolds with
Non-Abelian discrete symmetries.

The $\m A_5$ generators in its quartic representation are given by

\be
A_{\bf 4}\:= \pmatrix{
\hfill 1&\hfill 0&\hfill 0&\hfill 0&\cr 
\hfill 0&\hfill 1&\hfill 0&\hfill 0&\cr
\hfill 0&\hfill 0&\hfill -1&\hfill 0&\cr
\hfill 0&\hfill 0&\hfill 0&\hfill -1&} ;
\quad
B_{\bf 4}\:=\;\frac{1}{4}\pmatrix{ -1&-\sqrt {5}&-\sqrt {5}&-\sqrt {5}\cr \sqrt
  {5}&1&1&-3\cr -\sqrt {5}&-1&3&-1\cr\sqrt {5}&-3&1&1} . \ee
These matrices act on the real variables

\[\pmatrix{x^{}_1\cr x^{}_2\cr x^{}_3\cr x^{}_4}\ .\] From the Molien function of the four-dimensional representation \cite{Cummins:1988dr}

$$
\frac{1+\lambda^{10}}{(1-\lambda^2)(1-\lambda^3)(1-\lambda^4)(1-\lambda^5)} \ ,
$$
we see that there are four free and one constrained invariants.
The quadratic invariant is 

\be
I^{[2]}_{\bf 4}~=~(x^2_1+x^2_2+x^2_3+x^2_4)\ .
\ee
The cubic invariant is 

\be
I^{[3]}_{\bf 4}~=~-2\sqrt{5}x^{}_2x^{}_3x^{}_4-x^{}_1(x_2^2+x_3^2+x_4^2)+x_{1}^3\ .
\ee
The quartic invariant is

\be
I^{[4]}_{\bf 4}~=~5(x^4_2+x^4_3+x^4_4)-12\sqrt{5}x^{}_1x^{}_2x^{}_3x^{}_4+x^4_1 +12x^2_1(x^2_2+x^2_3+x^2_4)\ .
\ee
Hence the most general quartic invariant takes the form 

\be
I^{[4]}_{\m A_5}(\kappa)~=~I^{[4]}_{\bf 4}+\kappa \left(I^{[2]}_{\bf 4}\right)^2\ .
\ee
This invariant therefore defines a K3 surface which is left invariant under
the action of $\mathcal{PSL}_2(5)$.

 \section{Quintics from ${\boldsymbol{\mathcal A_6}}$}
This is the group of even permutations on six objects. It is simple, and its $360$ elements are products of  two generators  $c$ and $d$ with the presentation

\be<\,c,\,d\,|\,c^2=d^4=(cd)^5=(cd^2)^5=1\,>\ .
\ee
Its character table is 

\vskip .3cm
{\footnotesize{\begin{center}
\begin{tabular}{c|ccccccc}
&   \\
{\Large$ \m A_6$}~~~~& $C_1$&\hfill $45C^{[2]}_2$& \hfill$40C^{[3]}_3$&\hfill $40C^{[3]}_4$&\hfill $90C^{[4]}_5$&\hfill $72C^{[5]}_6$&\hfill $72C^{[5]}_7$\\
 &&&&&   \\
\hline    
 &&&&&   \\
$\chi^{[\bf 1]}_{}$~~& $1$&\hfill $1$&\hfill  $1$&\hfill  $1$&\hfill $1$&\hfill $1$&\hfill $1$\\
 &&&&&   \\
$\chi^{[\bf 5_1]}_{}$~~& $5$&\hfill  $1$&\hfill  $2$&\hfill  $-1$&\hfill $-1$&\hfill $0$&\hfill $0$ \\
 &&&&&   \\
$\chi^{[\bf 5_2]}_{}$~~& $5$&\hfill  $1$&\hfill  $-1$&\hfill  $2$&\hfill $-1$&\hfill $0$&\hfill $0$ \\
 &&&&&   \\
$\chi^{[\bf 8_1]}_{}$~~& $8$&\hfill  $0$&\hfill  $-1$&\hfill  $-1$&\hfill $0$&\hfill $-b^{}_5$&\hfill $-\tilde b^{}_5$\\
 &&&&&   \\
$\chi^{[\bf 8_2]}_{}$~~& $8$&\hfill  $0$&\hfill  $-1$&\hfill  $-1$&\hfill $0$&\hfill $-\tilde b^{}_5$&\hfill $ -b^{}_5$\\
 &&&&&   \\
$\chi^{[\bf 9]}_{}$~~& $9$&\hfill  $1$&\hfill  $0$&\hfill  $0$&\hfill $1$&\hfill $-1$&\hfill $-1$\\
 &&&&&   \\
$\chi^{[\bf 10]}_{}$~~& $10$& \hfill $-2$&\hfill  $1$&\hfill  $1$&\hfill $0$&\hfill $0$&\hfill $0$\\
\end{tabular}\end{center}
}}
\vskip .5cm

\noindent The ATLAS gives the form of the generators in the quintet representation
\be
c~=~ \pmatrix{
\hfill 1&\hfill 0&\hfill 0&\hfill 0&\hfill 0\cr 
\hfill 0&\hfill 0&\hfill 1&\hfill 0&\hfill 0\cr
\hfill 0&\hfill 1&\hfill 0&\hfill 0&\hfill 0\cr
\hfill 0&\hfill 0&\hfill 0&\hfill 0&\hfill 1\cr
\hfill 0&\hfill 0&\hfill 0&\hfill 1&\hfill 0} ;
\qquad
d~=~\pmatrix{
\hfill 0&\hfill 1&\hfill 0&\hfill 0&\hfill 0\cr
\hfill 0&\hfill 0&\hfill 1&\hfill 0&\hfill 0\cr
\hfill 0&\hfill 0&\hfill 0&\hfill 1&\hfill 0\cr
\hfill 1&\hfill 0&\hfill 0&\hfill 0&\hfill 0\cr
\hfill -1&\hfill -1&\hfill -1&\hfill -1&\hfill -1} .
\ee
The Molien function for the quintet representation can be easily computed. The result is

\[
\frac{1+\lambda^{15}}{(1-\lambda^2)(1-\lambda^3)(1-\lambda^4)(1-\lambda^5)(1-\lambda^6)  }\ ,
\]
which shows quadratic, cubic, quartic, quintic,  sextic and fifteenth-order invariants. The building of these invariants is facilitated by our knowledge of the  $\m A_5$  invariants. 

We first express the $\m A_6$ generators in the Cummins-Patera basis. We identify an $\m A_6$ element of order three, 

\[
b'~=~dcd^3(cd)^2 ~=~\pmatrix{
\hfill -1&\hfill -1&\hfill -1&\hfill -1&\hfill -1\cr
\hfill 0&\hfill 0&\hfill 0&\hfill 1&\hfill 0\cr
\hfill 1&\hfill 0&\hfill 0&\hfill 0&\hfill 0\cr
\hfill 0&\hfill 0&\hfill 0&\hfill 0&\hfill 1\cr
\hfill 0&\hfill 1&\hfill 0&\hfill 0&\hfill 0}\ ,\]
and fix the  similarity transformation by requiring

\be
A~=~\m U^{-1}_{} c\; \m U\ ,\qquad B~=~\m U^{-1}_{} b'\, \m U \ ,
\ee 
where $A$ and $B$ are the $\m A_5$ generators in the Cummins-Patera basis, see
Eq.~(\ref{AandB}). This yields

\[
\m U~=~\pmatrix{ \omega^2-1& 0& 0& \omega& -\omega\cr
 0& 0& 1-\omega^2& \omega^2& -1\cr
0&0&\omega^2-1&\omega^2&-1\cr
0&1-\omega^2&0&1&-\omega^2\cr
0&\omega^2-1&0&1&-\omega^2}\ .
\]
The $\m A_6$ order-four generator $d$ in the Cummins-Patera basis is given by

\be
d^{}_{CP}~=~\m U^{-1}_{} d\, \m U \ .
\ee
Explicitly

\be
d^{}_{CP}=~\frac{1}{2}\pmatrix{
0&-1&-1&1&1\cr
-2&0&0&0&0\cr
0&-1&-1&-1&-1\cr
0&\omega&-\omega&\omega&-\omega\cr
0&\omega^2&-\omega^2&-\omega^2&\omega^2}\ .
 \ee
Since $\m A_5 \subset \m A_6$, the most general $\m A_6$ quintic invariant can
be expressed in terms of the $\m A_5$ invariants of Section~\ref{A5sec} as

\[
  \alpha\,I^{[5]}_{s\,\bf 5}+\beta\, I^{[5]}_{a\,\bf 5}+\left(\gamma\,I^{[3]}_{s\,\bf 5}+\delta\,I^{[3]}_{a\,\bf 5}\right) I^{[2]}_{\,\bf 5}\ .
\]
The unknown coefficients $\alpha, \beta, \gamma, \delta$, are determined by requiring  that this expression be unaltered by the action of $d_{CP}$. This yields $\alpha=\gamma=0$; the quintic $\m A_6$ invariant 

\be
I^{[5]}_{\m A_{6}}(\kappa)~=~ I^{[5]}_{a\,\bf 5}+\kappa \,I^{[3]}_{a\,\bf 5}\,I^{[2]}_{\,\bf 5}\ ,
 \ee
depends only on one parameter. It is an invariant for both five-dimensional
irreps of $\m A_6$.

 \section{Quintics from ${\boldsymbol{\mathcal{PSL}_2(11)}}$}
This group has $660$ elements generated by two elements $C$ and $D$, with presentation
 
 \be
 <\,C,D\,|\,C^2=D^3=(CD)^{11}=[C,DCDCD]^2=1\,>\ .
\ee
Its character table is 
\vskip .3cm
{\footnotesize{\begin{center}
\begin{tabular}{c|ccccccccc}
&   \\
{\Large$ \mathcal{PSL}_2(11) $}~~~~& $C_1$&\hfill $55C^{[2]}_2$& \hfill$110C^{[3]}_3$&\hfill $132C^{[5]}_4$&\hfill $132C^{[5]}_5$
&\hfill $110C^{[6]}_6$&\hfill $60C^{[11]}_7$&\hfill $60C^{[11]}_8$ \\
 &&&&&&&&   \\
\hline    
 &&&&&&&&   \\
$\chi^{[\bf 1]}_{}$~~& $1$&\hfill $1$&\hfill $1$&\hfill $1$&\hfill $1$&\hfill $1$&\hfill $1$&\hfill $1$\\
 &&&&&&&&   \\
$\chi^{[\bf 5]}_{}$~~& $5$&\hfill  $1$&\hfill  $-1$&\hfill  $0$&\hfill $0$&\hfill $1$&\hfill $b_{11}^{}$&\hfill $\overline b_{11}^{}$ \\
 &&&&&&&&   \\
$\chi^{[\bf \overline 5]}_{}$~~& $5$&\hfill  $1$&\hfill  $-1$&\hfill  $0$&\hfill $0$&\hfill $1$&\hfill $\overline b^{}_{11}$&\hfill $b_{11}^{}$\\
 &&&&&&&&   \\
$\chi^{[\bf 10]}_{}$~~& $10$&\hfill  $-2$&\hfill  $1$&\hfill  $0$&\hfill $0$&\hfill$1$&\hfill$-1$&\hfill$-1$\\
 &&&&&&&&   \\
$\chi^{[\bf 10']}_{}$~~& $10$&\hfill  $2$&\hfill  $1$&\hfill  $0$&\hfill $0$&\hfill$-1$&\hfill$-1$&\hfill$-1$\\
 &&&&&&&&   \\$\chi^{[\bf 11]}_{}$~~& $11$&\hfill  $-1$&\hfill  $-1$&\hfill  $1$&\hfill $1$&\hfill$-1$&\hfill$0$&\hfill$0$\\
 &&&&&&&&   \\$\chi^{[\bf 12]}_{}$~~& $12$&\hfill  $0$&\hfill  $0$&\hfill  $b^{}_5$&\hfill $\tilde b_5^{}$&\hfill$0$&\hfill$1$&\hfill$1$\\
 &&&&&&&&   \\$\chi^{[\bf 12']}_{}$~~& $12$&\hfill  $0$&\hfill  $0$&\hfill  $\tilde b^{}_5$&\hfill $b_5^{}$&\hfill$0$&\hfill$1$&\hfill$1$\\
\end{tabular}\end{center}
}}
\vskip .3cm
\noindent where
\bea
 b^{}_{11}&=&\frac{1}{2}(-1+i\sqrt{11})~\:=~\:\eta+\eta^3+\eta^4+\eta^5+\eta^9\ ,\nonumber\\ \overline b^{}_{11}&=&\frac{1}{2}(-1-i\sqrt{11})~\:=\:~\eta^{10}+\eta^8+\eta^7+\eta^6+\eta^2\ ,
\eea
$\eta$ being  the eleventh root of unity, $\eta^{11}=1$.

The ATLAS expresses the generators as $(5\times 5)$ matrices
 
 \be
 C\,=\pmatrix{ 0&1&0&0&0\cr1&0&0&0&0
\cr0&0&0&0&1\cr1&-1&b^{}_{11}&1&-b^{}_{11}\cr0&0&1&0&0
}
 ;\quad
D\,=\pmatrix{ 0&0&0&1&0\cr 0&0&1&0&0
\cr 0&-1&-1&0&0\cr- \overline b^{}_{11}&0&0&-1&2+b^{}_{11}\cr1&0&0&-1&1
} ,
 \ee
acting on five complex numbers $(z^{}_1,z^{}_2,z^{}_3,z^{}_4,z^{}_5)$, the
components of its complex quintet irrep. In this basis the order-eleven
element $H$ is not diagonal,\\[-2mm] 
 \bea
H~=~ CD~=~\pmatrix{ 0&0&1&0&0\cr0&0&0&1&0
\cr1&0&0&-1&1\cr1&-b^{}_{11}&\overline b^{}_{11}&b^{}_{11}&2\cr 0
&-1&-1&0&0}\  .
 \eea
We take $C$ to be the  order-two $\m A^{}_5$
generator. $\mathcal{PSL}_2(11)$ has $110$ order-three generators, twenty of
which are $\m A^{}_5$ elements. We can choose the second  $\m A^{}_5$ generator to be 

\[
B'~=~DH^2D\ ,\qquad \Tr\, B'~=~-1\ ,\]
since it satisfies $\m A_5$'s presentation 

\[
 (B')^3~=~1\ ,\qquad (CB')^5~=~1\ .
\]
Explicitly, 

\[
B'~=~\pmatrix{ -b^{}_{11}&-2&b^{}_{11}-1&2&2\overline b^{}_{11}
\cr 1&\overline b^{}_{11}&-2&-\overline b^{}_{11}&
1-b^{}_{11}\cr 0
&0&1&0&0\cr 1&2&1-b^{}_{11}&-1+b^{}_{11}&2
\cr-1&2&1-b^{}_{11}&-1&b^{}_{11}}\  . \]
The next step is to find the similarity transformation $\m S$ which expresses these 
generators in the Cummins-Patera basis\\[-3mm]
\bea
A~=~\m S\, C\,\m S^{-1}_{}\ ,\qquad B~=~\m S\, B'\,\m S^{-1}_{}\ ,
\eea
where $A$ and $B$ were defined in Eq.~(\ref{AandB}). We find \\[-3mm]

\[\small
\m S~=~
\left(
\begin{array}{ccccc}
 0 & 2 (\omega+2) & -(b^{}_{11}-1) (\omega+2) & -2 (\omega+2) &
 -\overline b^{}_{11} (\omega+2) \\ 
 -2 (\omega+2) & 2 (\omega+2) & -b^{}_{11} (\omega+2) & 0 & b^{}_{11} (\omega+2) \\
 0 & 0 & -\omega-2 & 0 & \omega+2 \\
 -2 \overline b^{}_{11} & 6 & -3 b^{}_{11}-\omega+1 & 2 (b^{}_{11}-2) & 3
 b^{}_{11}-\omega+7 \\ 
 2 \overline b^{}_{11}\, \omega^2 & -6 \,\omega^2 & 2 \omega+3 b^{}_{11}
 \,\omega^2+1 & -2 (b^{}_{11}-2)\,\omega^2 & 8 \,\omega-3 b^{}_{11}\,\omega^2+7 
\end{array}
\right) .
\]

\noindent This allows us to calculate the order-three $\mathcal{PSL}_2(11)$
generator in the Cummins-Patera basis:    

\be
D^{}_{CP}~=~\m S\, D\,\m S^{-1}_{}\ ,
\ee
giving\\[-3mm]
\bea 
&&\hspace{-7mm}D^{}_{CP}~=~\frac{1}{12} \times \\[1mm]    
&&\hspace{-7mm}{\footnotesize{\pmatrix{3b^{}_{11}&3+3b^{}_{11}&3&-3-3b^{}_{11}\omega&-3-3b^{}_{11}\omega^2\cr 
-3& -6-3b^{}_{11}&-3-3b^{}_{11}&3+3b^{}_{11}+3\omega&3(b^{}_{11}-\omega)\cr
9+3b^{}_{11}&-3&-3b^{}_{11}&b_{11}^{}(\omega-1)+3 \omega &-b^{}_{11}(1+b^{}_{11}\omega^2)\cr
b^{}_{11}(\omega+2)-2\omega-1&5\omega+4+b^{}_{11}-b^{}_{11}\omega& -b^{}_{11}-5\omega-7-2b^{}_{11}\omega&-3+b^{}_{11}(2+\omega)&-2b^{}_{11}(1+2\omega)\cr
1+2\omega+b^{}_{11}(1-\omega)&b^{}_{11}(2+\omega)-1-5\omega&b^{}_{11}(2\omega+1)-2+5\omega&2b^{}_{11}(1+2\omega)&-3-b^{}_{11}(\omega-1)}
.}} \nonumber
\eea
 
\newpage

\subsection{Quintic Invariant} 
The Kronecker products of the  eight irreducible representations of
$\mathcal{PSL}_2(11)$ can be found in an unpublished note by the late
B. Wybourne~\cite{Wybourne}. They can be used to show the existence of {\em one} quintic invariant built out of its five-dimensional representation. The product  
 
\[
 {\bf 10'}\otimes{\bf 10'}=({\bf 1}+{\bf 5}+{\bf\overline 5}+{\bf 10'}+{\bf 10'}+{\bf 12}+{\bf 12'})_s+({\bf 10}+{\bf 11}+{\bf 12}+{\bf 12'})_a\ ,
 \]
 shows the existence of the invariant coupling  
 
 \[
 ({\bf 10'}\cdot{\bf 10'}\cdot{\bf 5})\ .\] 
 The second Kronecker product  
 
 \[
 {\bf 5}\otimes{\bf 5}~=~({\bf\overline 5}+{\bf 10'})_s+{\bf 10}_a\ , 
 \]
 allows us to decompose it in terms of quintets 
  
 \[
 \underbrace{\underbrace{{\bf 5}\otimes{\bf 5}}_{\bf 10'}\otimes\underbrace{ {\bf 5}\otimes{\bf 5}}_{\bf 10'}}_{\bf\overline 5}\otimes {\bf 5}\ .
 \]
Finally,   
 
 \[
 {\bf 5}\otimes{\bf \overline 5}~=~{\bf 1}+{\bf 12}+{\bf 12'}\ ,
 \]
leads to the quintic invariant built out of one quintet. Examination of the
other Kronecker products shows that it is unique.\footnote{Even though there
  exists a cubic invariant $I^{[3]}_{\m P \m S \m L_2(11)} =\,
  3\,(1+2\omega)\,I^{[3]}_{s\,{\bf 5}} + (3-2b_{11})\,I^{[3]}_{a\,{\bf 5}}$, no
  quintic invariant can be built from this due to the absence of a
  quadratic invariant.} This procedure is more 
convenient than computing the Molien function for this representation. 
 
 Consider the most general $\mathcal{PSL}^{}_2(5)$ quintic invariant  

\be
 I^{[5]}_{\mathcal{PSL}_2(11)}~=~ \alpha\,I^{[5]}_{s\,\bf 5}+\beta\,
 I^{[5]}_{a\,\bf 5}+\left(\gamma\,I^{[3]}_{s\,\bf 5}+\delta\,I^{[3]}_{a\,\bf
     5}\right)I^{[2]}_{\,\bf 5} \ , \label{PSL211inv}
\ee
written in terms of the five real variables $x^{}_i$ which span its real $\bf 5$ irreducible representation. However, in the embedding

\[
\mathcal{PSL}^{}_2(11)~\supset~ \mathcal{PSL}^{}_2(5)\ ,
\]
the quintets of the group and its subgroup match, that is

 \[
 {\bf 5}~=~{\bf 5}\ ,\qquad {\bf\overline 5}~=~{\bf 5}\ .
 \]
This allows us to replace in the invariant the real $x_i$ with the complex $z_i$ which span the complex $\bf 5$ of $\mathcal{PSL}^{}_2(11)$.

It is now a matter of acting $D^{}_{CP}$ on $ I^{[5]}_{\mathcal{PSL}_2(11)}$, and determine the coefficients in Eq.~(\ref{PSL211inv}) from invariance. This is an algebraically demanding exercise, which involves a combination of numerical and analytical tricks and the use of symbolic mathematical software.  However, a tedious calculation yields a simple answer
\bea
&&\alpha=1+13\,b^{}_{11}\ ,\quad~~ \beta=3(1+2\omega)(1-9b^{}_{11})\ ,\nonumber\\[1mm] &&\gamma=-8(2-b^{}_{11})\ ,\quad \delta=-8(1+2\omega)b^{}_{11}\ ,
\eea
which gives the $\mathcal{PSL}_2(11)$ quintic invariant in the Cummins-Patera basis.

\subsection{ A Convenient Basis}
This invariant can be considerably simplified when expressed in the basis where the order-eleven element is diagonal. If we perform the following similarity transformation on the generators

\be
\widetilde C~=~\m T^{-1}_{}\,C\,\m T\ ,\qquad \widetilde D~=~\m T^{-1}_{}\,D\,\m T\ ,
\ee
where 

\[
\m T=
\pmatrix{ 
{\eta}^{4}+{\eta}^{6}  &  {\eta}^{}+{\eta}^{7} & {\eta}^{2}+{\eta}^{5} & {\eta}^{8}+{\eta}^{9} & {\eta}^{3}+{\eta}^{10} \cr
-\eta^{7}&-\eta^{10}&-\eta^{6}&-\eta^{2}&-\eta^{8}\cr
 {\eta}^{5}+\eta^7 &\eta^4+{\eta}^{10} &\eta^6+{\eta}^{9} &\eta^2+{\eta}^3& {\eta}^{}+\eta^8 \cr
-\eta^8&-{\eta}^{2}&-{\eta}^{10}&\,-{\eta}^{7}&-{\eta}^{6}\cr
-{\eta}^{4}&  -{\eta}^{}&-{\eta}^{5}& -{\eta}^{9}&  -{\eta}^{3}}\ ,\qquad \det\m T=11\ ,
\]
we find that 

\be
\widetilde{  H} ~=~ \widetilde C\,\widetilde D \ ,
\ee
becomes  diagonal

\be
\widetilde { H}~=~\mbox {Diag}(\eta,\eta^3,\eta^4,\eta^5,\eta^9)\ .
 \ee
$\widetilde { H}$ generates  the $\m Z_{11}$ subgroup. The second order-eleven class of $\mathcal{PSL}_2(11)$ is generated by its conjugate.  

The order-two generator $\widetilde C$ is real. Applying the notation
$$
c_k ~=~ \cos \left( \frac{2 \pi k}{11} \right) \ ,
$$
its explicit form reads
\begin{eqnarray}
&&\hspace{-7mm}\widetilde C~=~\frac{1}{11}\times  \\[1mm]
&&\hspace{-7mm}{\tiny{\left(\!\!\!\!\!
\begin{array}{ccccc}
 -4 c_1-2 c_2+4 c_5+2 \!&\! 6 c_2+2 c_4-4 c_5-4 \!&\! 2 c_1+10 c_3+8 c_4+2 \!&\! 2 c_1+10 c_3+8 c_4+2 \!&\! -10 c_2-6 c_3-12 c_4-5 \\[1mm]
 8 c_1+10 c_2+2 c_3+2 \!&\! -4 c_3+4 c_4-2 c_5+2 \!&\! 8 c_1+10 c_2+2 c_3+2 \!&\! -12 c_1-6 c_2-10 c_5-5 \!&\! 2 c_1-4 c_4+6 c_5-4 \\[1mm]
 -4 c_2+6 c_3+2 c_5-4 \!&\! -6 c_1-10 c_3-12 c_5-5 \!&\! 4 c_2-2 c_3-4 c_4+2 \!&\! 10 c_1+2 c_4+8 c_5+2 \!&\! 10 c_1+2 c_4+8 c_5+2 \\[1mm]
 -10 c_1-12 c_2-6 c_4-5 \!&\! 8 c_2+10 c_4+2 c_5+2 \!&\! 6 c_1+2 c_2-4 c_3-4 \!&\! -2 c_1+4 c_3-4 c_5+2 \!&\! 8 c_2+10 c_4+2 c_5+2 \\[1mm]
 2 c_2+8 c_3+10 c_5+2 \!&\! 2 c_2+8 c_3+10 c_5+2 \!&\! -12 c_3-10 c_4-6 c_5-5 \!&\! -4 c_1+2 c_3+6 c_4-4 \!&\! 4 c_1-4 c_2-2 c_4+2
\end{array}\!\!\!
\right).}} \nonumber
\end{eqnarray}
Since $\widetilde C$ is an involution, the second $\mathcal{PSL}_2(11)$ generator is easily obtained from 

\be
\widetilde D~=~\widetilde C\,\widetilde{ H}\ .
\ee
The subgroup $\m A_5$ is now generated by the order-two and order-three
elements 

$$
\widetilde C \ , \qquad \widetilde {B}' ~=~\widetilde D \widetilde H^2
\widetilde D \ .
$$

It is known since Galois, see e.g. Ref.~\cite{Kostant}, that every element $g
\in \mathcal{PSL}_2(11)$  can be written uniquely as 
 
 \[g=k\cdot h\ ,
 \]
 where $k\in \m A_5$ and $h\in \m Z_{11}$. In this basis, where $h$ is diagonal, it is easy to determine the  eleven monomials which are invariant under $\m Z_{11}$. Written in terms of the components of the quintet $z^{}_i$, $i=1,2,\dots,5$, they are

 $$
 \prod_{i=1}^5\,z_i^{a_i}\ ,
 $$
 with
 
 $$
 a^{}_1+3a^{}_2+4a^{}_3+5a^{}_4+9a^{}_5~=~0 ~ {\rm mod}\,(11)\ .$$
For the fifth-order monomials that is
\begin{eqnarray*}
&  z_1^3\,z_3^2, ~ \, z_5^3\,z_2^2,  ~ \,z_3^3\,z_4^2 ,~\,z_2^3\,z_1^2,    ~\,
z_4^3\,z_5^2 \ , 
 \\
&  z_1^3\,z_2^{}\,z_4^{}, 
~\,z_5^3\,z_4^{}\,z_1^{}, 
~\, z_3^3\,z_1^{}\,z_5^{},
~\, z_2^3\,z_5^{}\,z_3^{},
~\, z_4^3\,z_3^{}\,z_2^{} \ , 
\\
& z_1z_2z_3z_4z_5\ .
\end{eqnarray*}
The $\m A_5$ elements $\widetilde C$ and  $\widetilde {B}'$ then rotate these monomials into one another so as to achieve invariance. The quintic invariant is a linear combination of these eleven monomials. A laborious calculation yields

\begin{eqnarray}
I^{[5]}_{\m P\m S\m L_2(11)} 
&\!\!\!\!\!=\!\!\!& z_1^3 z_3^2 \left(-5-2 \eta +6 \eta ^2+10 \eta ^3+12 \eta
  ^4+15 \eta ^5+10 \eta ^6+3 \eta ^7-6 \eta ^9-8 \eta ^{10} \right)\nonumber\\
&& \!\!\!\!\!\!\!\:\,+ z_5^3 z_2^2 \left(-5+15 \eta -8 \eta ^2+12 \eta ^3-6 \eta ^4+10 \eta ^5+6 \eta ^7+3 \eta ^8-2 \eta ^9 + 10 \eta ^{10} \right)\nonumber\\
&&\!\!\!\!\!\!\!\:\,+ z_3^3 z_4^2 \left( -5+10 \eta+10 \eta ^2-6 \eta ^3-2
  \eta ^4+12 \eta ^5+3 \eta^6-8 \eta ^7+6 \eta ^8+15 \eta ^9 \right)\nonumber\\
&&\!\!\!\!\!\!\!\:\,+ z_2^3 z_1^2 \left(-5+12 \eta -2 \eta ^3+15 \eta ^4-6
  \eta ^5+6   \eta ^6+10 \eta ^7-8 \eta ^8+10 \eta ^9+3 \eta ^{10} \right)\nonumber\\
&&\!\!\!\!\!\!\!\:\,+z_4^3 z_5^2 \left(-5-6 \eta +3 \eta ^2+15 \eta ^3+10 \eta
  ^4-2 \eta ^5-8 \eta ^6+10 \eta ^8+12 \eta ^9 +6 \eta ^{10} \right)\nonumber\\
&&\!\!\!\!\!\!\!\:\,+z_1^3 z^{}_2 z^{}_4\left(-1-3 \eta +2 \eta^2+3 \eta^3+3 \eta
  ^4+5 \eta^5+3 \eta ^6-4 \eta ^9-4 \eta ^{10}\right)\nonumber\\
&&\!\!\!\!\!\!\!\:\,+ z_5^3 z^{}_4 z^{}_1 \left(-1+5 \eta -4 \eta ^2+3 \eta ^3-4
  \eta ^4+3 \eta^5+2 \eta ^7-3 \eta ^9 +3 \eta ^{10}\right)\nonumber\\
&&\!\!\!\!\!\!\!\:\,+ z_3^3 z^{}_1 z^{}_5 \left( -1+3 \eta+3 \eta ^2-4 \eta ^3-3
  \eta ^4+3 \eta^5-4 \eta ^7+2 \eta ^8 +5 \eta ^9\right)\nonumber\\
&&\!\!\!\!\!\!\!\:\,+z_2^3 z^{}_5 z^{}_3 \left(-1+3 \eta -3 \eta ^3+5 \eta ^4-4 \eta ^5+2 \eta^6+3 \eta ^7-4 \eta ^8+ 3 \eta ^9\right)\nonumber\\
&&\!\!\!\!\!\!\!\:\,+ z_4^3 z^{}_3 z^{}_2 \left(-1-4 \eta +5 \eta ^3+3 \eta ^4-3
  \eta ^5-4 \eta^6+3 \eta ^8+3 \eta ^9+2 \eta ^{10}\right)\nonumber\\
&&\!\!\!\!\!\!\!\:\,-3 z^{}_1 z^{}_2 z^{}_3 z^{}_4 z^{}_5 \left(-2+\eta +\eta ^3+\eta ^4+\eta^5 + \eta ^9 \right) .\label{PSLinv}
\end{eqnarray}

\vspace{2mm}

In order to understand this expression a little bit better, note that a 
particular conjugation of the order-five generator $\widetilde C \widetilde
{B}'$ takes a rather simple form:\\[-3mm]
\begin{eqnarray*}
&&\hspace{-7mm} \left(\widetilde B' \widetilde{ H}^4 \right)^{-1}_{} 
   \left(\widetilde C \widetilde B' \right) 
   \left(\widetilde B' \widetilde{ H}^4 \right)~=~ \\[1mm]
&&\hspace{-7mm}{\footnotesize{
\left(\!\!\!\begin{array}{ccccc} 
0&0&0&0& 1+\eta^2+\eta^6+\eta^7 \\ 
0&0&0&1+\eta^6+\eta^7+\eta^{10}&0 \\
0&1+\eta^2+\eta^6+\eta^{8} &0&0&0 \\
1+\eta^2+\eta^8+\eta^{10} &0&0&0&0 \\
0&0&1+\eta^7+\eta^8+\eta^{10}&0&0  \end{array}\!\!\right)  .}}
\end{eqnarray*}
Its action on the complex variables $z_i$ relates the first five rows
of Eq.~(\ref{PSLinv}) among each other as well as the next five rows. For
instance, replacing 
\begin{eqnarray*}
z_1 &\rightarrow & z_1'\,=\,z_5 \,(1+\eta^2+\eta^6+\eta^7) \ ,\\
z_3 & \rightarrow & z_3'\,=\,z_2 \,(1+\eta^2+\eta^6+\eta^8) \ ,
\end{eqnarray*}
in the first line of Eq.~(\ref{PSLinv}) yields the second one.

Of course, the quintic invariant can be written in many different ways. 
Arranging the monomials in the diagonal matrices 
\bea
\m E&=&{\rm Diag}\,( z_5^3\,z_2^2, \, z_4^3\,z_5^2,  \,z_2^3\,z_1^2,
\,z_1^3\,z_3^2, \,z_3^3\,z_4^2\,) \ , \nonumber \\[1mm]
\m F&=&{\rm Diag}\,(z_5^3\,z_4^{}\,z_1^{}, \,z_4^3\,z_3^{}\,z_2^{}, \,z_2^3\,z_5^{}\,z_3^{}, \,z_1^3\,z_2^{}\,z_4^{}, \,z_3^3\,z_1^{}\,z_5^{}\,)\ ,
\eea
the quintic invariant $I^{[5]}_{\mathcal{PSL}_2(11)}$ of Eq.~(\ref{PSLinv}) can be rewritten as 

\bea
&&\hspace{-7mm}I^{[5]}_{\mathcal{PSL}_2(11)}~=~3\,z_1z_2z_3z_4z_5(2-b^{}_{11})~ \nonumber\\[1mm]
&&\hspace{-7mm}+~\Tr\left(\m E\,[\,15\widetilde { H}-8\widetilde { H}^2
+12\widetilde { H}^3-6\widetilde { H}^4+10\widetilde { H}^5+6\widetilde {H}^7
 +3\widetilde { H}^8-2\widetilde { H}^9+10\widetilde { H}^{10}-5\,]\right) 
\nonumber\\[1mm]
&&\hspace{-7mm}+~\Tr\left(\m F\,[\,5\widetilde { H}-4\widetilde { H}^2
+3\widetilde { H}^3-4\widetilde { H}^4+3\widetilde { H}^5+2\widetilde { H}^7
-3\widetilde { H}^9+3\widetilde { H}^{10}-1\,]\,\right) .
\eea
Setting this invariant to zero determines a unique Calabi-Yau three-fold
with the largest simple Non-Abelian finite symmetry group. Its group,
$\mathcal{PSL}_2(11)$, encodes the geometry of the truncated icosahedron~\cite{Kostant,Kostant2} (football, $C_{60}$, quasicrystals), which has twelve pentagonal faces, and $20$ hexagonal faces. Its detailed study should be of some interest to both physicists and mathematicians. 

\section{No Quintics from ${\boldsymbol{U_4(2)= Sp_4(3)}}$}
The simple group of $(4\times 4)$ unitary matrices over the Galois field
of order two contains 25920 elements, and has a complex five-dimensional
representation as well as its conjugate. 
It is generated by two elements $a$ and $b$ with presentation
 
 \be
 <a,b\,\vert \, a^2=b^5=(ab)^9=[\,a,b\,]^3=[\,a,bab\,]^2=1> \ .
 \ee
 In the quintet, the ATLAS shows that they are given by
 
 \be
 a\:=\pmatrix{ -1&\hfill 0&\hfill 0&0&0\cr \hfill 0&-1&\hfill 0&0&0 \cr \hfill
   0&\hfill 0&-1&0&0\cr \hfill 0&\hfill 0&\hfill 0&0&1\cr \hfill 0&\hfill
   0&\hfill 0&1&0},\ \quad
 b\:=\pmatrix{  0& 1& 0&0&0\cr  0& 0& 1&0&0 \cr  0& 0& 0&1&0\cr  \omega^2& 1&
   -\omega^2&\omega&-\omega\cr  0& -\omega^2& 0&0&-\omega}. 
 \ee
This group has twenty irreducible representations. In order to find out the degree of the invariants made out of  quintets, we need to compute the Molien function. For that we need a representative from each class.
 
 \begin{itemize}
 
\item One class of order one.
\item Two classes of order two:
 
 \begin{enumerate}
 \item  generated by $a$, with trace $=-3$\,;
 \item  generated by $[\,a,bab\,]$, with trace $=1$\,.
 \end{enumerate}
  \item Four classes of order three:
 
 \begin{enumerate}
\item  generated by $[\,a,b\,]$, with trace $=2$\,;
\item  generated by $b^2ab[\,a,bab\,]$, with trace $=-\overline b^{}_{27}$\,;
\item  generated by $a[\,a,bab\,][\,a,b\,]b^2$, with trace $=- b^{}_{27}$\,;
\item  generated by $b^4ab^3ab^4ab^3a$, with trace $=-1$\,.
 \end{enumerate}

  \item Two classes of order four:
 \begin{enumerate}
\item generated by $ab^3ab^3ab^3$, with trace $=1$\,;
\item generated by $a[\,a,bab\,][\,a,b\,]$, with trace $=2$\,.
 \end{enumerate} 
  
\item One class of order five: generated by $b$, with trace $=0$\,.

\item Six classes of order six:
\begin{enumerate}
\item generated by $ab^2ab^2$, with trace $=\overline \omega-1$\,;
\item generated by $b^2a[\,a,bab\,][\,a,b\,]a$, with trace $=\omega-1$\,;  
 \item generated by $b^2[\,a,b\,]$, with trace $=i\sqrt{3}$\,;
\item generated by $babab[\,a,b\,]$, with trace $=-i\sqrt{3}$\,;
 \item generated by $b^2ab^3ab^3[\,a,b\,][\,a,bab\,]$, with trace $=0$\,;
\item generated by $ab^4[\,a,bab\,][\,a,b\,]$, with trace $=1$\,.
\end{enumerate} 
  
 \item Two classes of order nine:
 \begin{enumerate}
 \item  generated by $ab$, with trace $=-\omega$\,;
 \item  generated by $ab^4$, with trace $=-\overline\omega$\,.
 \end{enumerate}
  
 \item Two classes of order twelve:
 \begin{enumerate}
 \item  generated by $b[\,a,b\,]$, with trace $=\omega$\,;
 \item   generated by $bab$, with trace $=\overline\omega$\,.
 \end{enumerate}
 \end{itemize}
Using the method outlined in Section~\ref{molien} we compute the Molien generating function for the invariants constructed out of one quintic

$$F({\bf 1};{\bf 5};\lambda)~=~\frac{1}{25920}\sum _{i=1}^{20}\,\frac{n_i^{}}{\det(1-\lambda A^{[{\bf 5}]}_i)}\ ,$$
where $A^{[{\bf 5}]}_i$ is any element in the $i$th class.  The result from MAPLE is

{\small{$$
{\frac { \left( {\lambda}^{24}+{\lambda}^{21}-{\lambda}^{15}-{\lambda}^{12}-{\lambda}^{9}+{\lambda}^{3}+1
 \right)  \left( {\lambda}^{8}+{\lambda}^{7}-{\lambda}^{5}-{\lambda}^{4}-{\lambda}^{3}+\lambda+1 \right) }
{ \left( {\lambda}^{4}+{\lambda}^{3}+{\lambda}^{2}+\lambda+1 \right)  \left( {\lambda}^{8}+{\lambda}^{4}+1
 \right)  \left( {\lambda}^{6}+{\lambda}^{3}+1 \right)  \left( 1+{\lambda}^{2}+{\lambda}^{4}
 \right)  \left( {\lambda}^{2}+\lambda+1 \right)  \left( \lambda+1 \right) ^{4} \left( {
\lambda}^{2}+1 \right) ^{2} \left( \lambda-1 \right) ^{5}}}\ .$$}}

\noindent This function can be written in a suggestive form with  the
denominator as a product of functions of the form $(1-\lambda^n)$, and the
numerator as a finite polynomial with positive coefficients, that is

\[
F({\bf 1};{\bf
  5};\lambda)~=~\frac{(1+\lambda^{10}+\lambda^{20})(1+\lambda^{30}+\lambda^{60})}{(1-\lambda^4)(1-\lambda^6)(1-\lambda^{12})(1-\lambda^{18})(1-\lambda^{45})}
\ .
\]
It clearly contains no term of order $\lambda^5$, showing that there is no quintic invariant. 
  
\section{Acknowledgments}
One of us (PR) would like to acknowledge  the Ambrose Monell foundation for
support, and the Institute for Advanced Study where the early part of this
work was done, as well as informative discussions with Drs Johannes  Walcher,
Yuji Tachikawa, and Simon Judes.  Both of us wish to specially thank Dr Johannes Walcher for cross-checking some of
our results and also Prof John McKay for his many remarks on Molien functions. PR is supported in part by the Department Of Energy Grant No. DE-FG02-97ER41029. CL is supported by the University of Florida through the Institute for Fundamental Theory.


\end{document}